\documentstyle{article}
\begin{document}
\title
 {High Velocity Neutron Stars as a Result of Asymmetric Neutrino
 Emission
 \thanks{This work was supported in part by RFFI grant 93-02-17106,
 Astronomy Programm of RMSTP topic 3-169, and KOSMION}}
\author{G.S.Bisnovatyi-Kogan}
\date{}
\maketitle

{\centerline{\it Space Recearch Institute}}

{\centerline{\it Profsoyuznaya 84/32 Moscow 117810, Russia}}

\bigskip

{\bf AIP CONFERENCE PROCEEDINGS 366}

{\bf High Velocity Neutron Stars and Gamma-Ray Bursts}

La Jolla, CA 1995; pp.38-42.

\medskip

EDITORS

Richard Rothschild

Richard E. Lingenfelter

\bigskip

\begin{abstract}
  Formation of a neutron star is accompanied by
neutrino emission carring about 10\% of the rest energy of the star.
Toroidal field produced
by twisting of a dipole field in differentially rotating
star is antisymmetric. Its summation with antisymmetric
toroidal field results in braking of mirror symmetry of the
magnetic field .
 For large magnetic field the neutron decay rate depends on its
 strength.
Neutrino is emitted more in one direction leading to
flux asymmetry and recoil of the neutron star. Estimations show that
the
neutron star can reach velocities $\sim 1000$ km/s for $3\%$ asymmetry
of
the neutrino flux.
\end{abstract}

\section*{Introduction}
Observations of the pulsars moving at the velocities up to 500 km/s
\cite{Harrison} is a
challenge to the theory of the neutron star formation. The
Blaauw effect during the formation of pulsars
in the binaries cannot produce such a high speeds. The plausible
explanation for the birth of rapidly mooving pulsars seems to be the
suggestion of the kick at the birth from the asymmetric explosion
\cite{Shklovskii}. We make estimations
for the strength of the kick, produced by the asymmetric neutrino
emission during the collapse.

The asymmetry of the neutrino pulse, is produced by the
asymmetry of the magnetic field distribution, formed during the
collapse and
differential rotation \cite{BisnovatyiMoiseenko}.
\cite{BisnovatyiKogan1993}.

\section*
{Formation of the asymmetric magnetic field}
Consider rapidly and differentially rotating new born neutron star
with the dipole poloidal and symmetric toroidal fields.
A field amplification during the differential rotation leads to
the formation of additional
toroidal field from the poloidal one. This field, made from the
dipole poloidal one by twisting is antisymmetric
with respect to the symmetry plane. The sum of the initial symmetric
with
the induced antisymmetric toroidal fields has no plane symmetry.

In absence of
dissipative processes
the neutron star returns to the state of rigid rotation loosing the
induced toroidal field and restoring mirror symmetry of the matter
distribution.  Formation of asymmetric toroidal field distribution
is followed by magnetorotational explosion
\cite{BisnovatyiKogan1970}, which is asymmetric, leading
to neutron star recoil and star acceleration
\cite{BisnovatyiMoiseenko}.
The neutron star accelleration happens also due to dependence
of the cross-section of week interactions on the magnetic field.

The influence of the magnetic field on the
neutron decay was studied in \cite{OConnel}. It becomes essential
when the characteristic energy of the electron at the
Landau level with the
Larmor rotation ${\hbar eB \over m_e c}$ becomes of the order of the
energy
of the decay $\sim m_e c^2$ for a neutron. That happens at
 \begin{equation}
 B_c={m_e^2 c^3 \over e \hbar} = 4.4 \times 10^{13} {\rm Gs}
 \label{eq:s1}
 \end{equation}
\noindent
The probability of the neutron decay $ W_n $ in vacuum is
 $$
 W_n=W_0[1+0.17(B/B_c)^2+...] \quad {\rm at} \quad B \ll B_c
 $$
 \begin{equation}
 W_n=0.77 W_0 (B/B_c) \quad {\rm at } \quad B \gg B_c
 \label{eq:s2}
 \end{equation}
\noindent
At high temperatures or densities the critical field increases in
$\alpha_T=kT/m_e c^2$ or $\alpha_{\rho}=\epsilon_{Fe}/m_e c^2$ times.
After a collapse of rapidly rotating star the neutron star
rotates at the period $P$ about 1 ms.
Differential rotation leads to the linear
amplification of the toroidal field
 \begin{equation}
 B_{\phi}=B_{\phi 0}+B_p (t/P)
 \label{eq:s3}
 \end{equation}
The time of the neutrino emission is several tens of seconds
\cite{Nadjozhin}.
After 20 s the induced toroidal magnetic field
will be about $ 2 \times 10^4 B_p $, corresponding to $ 10^{15}
\div 10^{17} $ Gs for $ B_p = 10^{11} \div 10^{13} $ Gs, observed in
the
pulsars. Adopting the initial toroidal field
$B_{\phi 0}=(10 \div 10^3) B_p = 10^{12} \div 10^{16} $, we may
estimate an asymmetry of the neutrino pulse.
For symmetric $ B_{\phi,0} $
and dipole poloidal field the difference
$\Delta B_{\phi} $ between the magnetic
fields absolute values
in two hemispheres increases, until it reaches the value
$ 2 B_{\phi 0} $. It remains constant later, while the relative
difference
 \begin{equation}
\delta_B={\Delta B_{\phi} \over B_{\phi +} + B_{\phi -} }
 \label{eq:si}
 \end{equation}
\noindent
decreases.

\section*
 {Neutrino heat conductivity and energy losses }
The main neutrino flux is formed in the region where the mean free
path of
the neutrino is smaller then the stellar radius. The neutrino energy
flux,
$ H_{\nu} $ associated
with the temperature gradient may be written as \cite{Imshennik}
 \begin{equation}
 H_{\nu}=-{7 \over 8} {4acT^3 \over 3} l_T {\partial T \over
 \partial r}
 \label{eq:s4}
 \end{equation}
\noindent
The quantity $l_T$ having the meaning of the neutrino mean free path
is connected with the neutrino opacity $ \kappa_{\nu} $ as
 \begin{equation}
\kappa_{\nu}=1/(l_T \rho)
 \label{eq:s5}
 \end{equation}
\noindent
Calculations of the spherically symmetrical collapse
\cite{Nadjozhin} have
shown, that during the phase of the main neutrino emission, a hot
neutron star consists
of the quasiuniform quasiisothermal core with the temperature $T_i$,
whose mass increases with
time, and the region between the neutrinosphere and
the isothermal core, where the
temperature smoothly decreases in about 10 times while
the density, which finally drops about 6 times decreases
nonmonotonically.
Neutrino flux is forming in this
region, containing about one half of the neutron star mass.
We suggest for sumplicity  a power-law
dependences for the temperature and $ l_T $:
 \begin{equation}
 T=T_i \left( {r_i \over r} \right)^m, \qquad
 l_T={1 \over \kappa \rho }=l_{Ti} \left(
{r \over r_i} \right)^n
 \label{eq:s6}
 \end{equation}
\noindent
The neutrinosphere with the radius $r_{\nu} $
is determined approximately by the relation
 \begin{equation}
 \int_{r_{\nu}}^{\infty} \kappa_{\nu} \rho dr = \int_
{r_{\nu}}^{\infty} {dr \over l_T}=1
 \label{eq:s8}
 \end{equation}
\noindent
Using (~\ref{eq:s6}) outside the neutrinosphere we get
from (~\ref{eq:s8}) the relation
 \begin{equation}
 r_{\nu}=r_i{ \left( r_i \over (n-1)l_{Ti}
\right) }^{1 \over n-1}
 \label{eq:s9}
 \end{equation}
\noindent
From (~\ref{eq:s4})-(~\ref{eq:s6}), using (~\ref{eq:s9}) we get the
temperature of the neutrinosphere $T_{\nu}$
and the heat flux on this level $H_{\nu}$, which outside the
neutrinosphere
is approximately $\sim r^{-2}$, corresponding to the constant neutrino
luminosity $L_{\nu}$
 \begin{equation}
 T_{\nu}=T_i \left( (n-1)l_{Ti} \over r_i \right)^{m \over n-1}
 \label{eq:s10}
 \end{equation}
 \begin{equation}
 L_{\nu}=4\pi r_{\nu}^2 H_{\nu}={7 \over 8} m {16 \pi acT_i^4 \over 3}
(n-1)^{4m-n+1 \over n-1} r_i^2 \left(l_{Ti} \over r_i \right)
^{4m-2 \over n-1}
 \label{eq:s11}
 \end{equation}
\noindent
To estimate the anisotropy of the neutrino flux we compare
two stars with the same radius and temperature of the core $ r_i$
and $ T_i$ and different opacities.
Let $l_{Ti}$ is different
and constant in two hemispheres, and each one is radiating  according
to
(~\ref{eq:s11}). The anisotropy of the flux
 \begin{equation}
 \delta_L ={L_+-L_- \over L_++L_-}
 \label{eq:s12}
 \end{equation}
\noindent
here $L_+$, $L_-$ are luminosities in the different hemispheres,
calculated, using (~\ref{eq:s11}). For small difference between
hemispheres
 \begin{equation}
\delta_L={\Delta L \over L}={4m-2 \over n-1} {\Delta l_{Ti} \over
l_{Ti}}
 \label{eq:s13}
 \end{equation}
\noindent
Here $n>1$, when $m={1 \over 2} $ the neutrino
fluxes in both
hemispheres are equal because smaller opacity and larger neutrinosphere
temperature $T_{\nu}$ from (~\ref{eq:s10}) is compensated by
smaller neutrinosphere
radius $r_{\nu}$ from (~\ref{eq:s9}).

\section*
{Neutron star acceleration.}
The equation of motion of the neutron star with the
mass $ M_n $
\begin{equation}
M_n{dv_n \over dt }= {L_{+}-L_{-} \over c}, \qquad
L_{+}+L_{-}= {2 \over \pi} L_{\nu}(t)
 \label{eq:s17}
 \end{equation}
\noindent
For the power distributions (~\ref{eq:s9}),(~\ref{eq:s10}) it follows
from (~\ref{eq:s11}) that
 \begin{equation}
 L_{\pm}=A l_{Ti\pm}^{4m-2 \over n-1}
 \label{eq:s18}
 \end{equation}
\noindent
In general $l_{Ti}$ is determined by various neutrino processes and
depends on $B$.
As an example consider the dependence on $B$ in the form
(~\ref{eq:s2}). Making
interpolation between two asimptotic forms we get dependence
 \begin{equation}
 l_{Ti\pm} \sim {1 \over W} =l_{T0} {1+{ \left (B \over B_c \right )}^3
\over 1+0.17 {\left( B \over B_c \right) }^2+0.77{\left( B \over B_c
\right) }^4}=l_{T0}F^{n-1 \over 4m-2}(B)
 \label{eq:s19}
 \end{equation}
\noindent
The time dependence of the average value of $B$ in each hemisphere can
be
found from (~\ref{eq:s3}) with
 \begin{equation}
 B_{p+}=-B_{p-}, \quad  B_{\phi 0 +}=B_{\phi 0 -}
 \label{eq:s20}
 \end{equation}
\noindent
By $B_p$ we mean a radial component of the poloidal
field taking part in amplification of $B_{\phi}$.
The time dependence of $L_{\nu}$ is taken from the spherically
symmetric calculations of the collapse.

\section*
 {Quantitative estimations }
 For ${4m-2 \over n-1}=1 $ and in condition when
 the neutron star is accelerated at $B \gg B_c$, we have
$ F_{\pm}={B_c \over 0.77 B_{\pm} }$. Equation of motion
(~\ref{eq:s17}) may be written as
\begin{equation}
 M_n {dv_n \over dt}={2 \over \pi} {L_{\nu} \over c} {\vert B_+\vert -
\vert B_-\vert \over \vert B_+\vert +\vert B_-\vert }
 \label{eq:s26}
 \end{equation}
\noindent
with the linear functions for $B_{\pm}$.
Take constant $ L_{\nu}={0.1 M_n c^2 \over 20 s}  $.
With these simplifications, the final velocity of the neutron star
$v_{nf}$
follows as a result of the solution of (~\ref{eq:s26}) in the form
 \begin{equation}
 v_{nf}={2 \over \pi} {L_{\nu} \over M_n c}{PB_{\phi 0} \over \vert B_p
\vert} (0.5+\ln \left( {20 \, s
\over P} {\vert B_p \vert \over B_{\phi 0}} \right) )
 \label{eq:s29}
 \end{equation}
\noindent
For $P=10^{-3} \, s $ we obtain from (~\ref{eq:s29})
 \begin{equation}
 v_{nf}={2 \over \pi} {c \over 10} {P \over 20 \, s} x(0.5+
\ln({20 \, s \over P} {1 \over x} )) \approx 1 {km \over s} x(0.5+\ln
{2\times 10^4 \over x})
 \label{eq:s30}
 \end{equation}
\noindent
For the value $ x={B_{\phi 0} \over \vert B_p \vert} $  ranging
between $20$
and $10^3$, we have $v_{nf}$ between 140 and 3000 km/s, what can
explain
the nature of the most rapidly mooving pulsars.
The formula (~\ref{eq:s30}) can
be applied when
$ B_{\phi 0} \gg B_c $ and $ x \gg 1$.

The acceleration of the collapsing star by anisotripic neutrino
emission can
happen even when the star collapses to the black hole,
the efficiensy of acceleration decreases with increasing of mass.
 We may expect black holes of stellar
origin moving rapidly, like radiopulsars, and
they may be found high over the galactic disk.
This is observed among the soft X ray novae -
most probable candidates for black holes in
the Galaxy \cite{Greiner}.

After magnetorotational explosion we may expect tens of ms periods for
a neutron star rotation. Remaining in binary, star either accelerates
its rotation (in LMXB) or decelerates it (in high mass XB). The first
ones
are transformed into recycled (msec, binary) pulsars, and the last
ones
(after explosion of the massive component and disruption of binary)
may form a family of very slowly rotating neutron stars, one of which
was
observed in the strongest GRB of 5 March 1979.

\end{document}